\documentclass[%
 aip,
 amsmath,amssymb,
 reprint,%
]{revtex4-1}

\usepackage{graphicx}
\usepackage{dcolumn}
\usepackage{bm}

\usepackage[utf8]{inputenc}
\usepackage[T1]{fontenc}
\usepackage{mathptmx}
\usepackage{etoolbox}

\def\bm#1{\mbox{\boldmath{$#1$}}}
\def\rr#1{(\ref{#1})}

\newcommand{\be}{\begin{equation}}
\newcommand{\ee}{\end{equation}}

\makeatletter
\def\@email#1#2{%
 \endgroup
 \patchcmd{\titleblock@produce}
  {\frontmatter@RRAPformat}
  {\frontmatter@RRAPformat{\produce@RRAP{*#1\href{mailto:#2}{#2}}}\frontmatter@RRAPformat}
  {}{}
}%
\makeatother
\begin{document}

\preprint{AIP/123-QED}

\title[]{Taylor halos and Taylor spears in odd viscous liquids}
\author{E. Kirkinis}
\affiliation{ 
Department of Materials Science \& Engineering, Northwestern University, Evanston IL 60208 USA
}%
\affiliation{ 
Center for Computation and Theory of Soft Materials, Northwestern University, Evanston IL 60208 USA
}%
 \email{kirkinis@northwestern.edu}
\author{M. Olvera de la Cruz}%

\affiliation{ 
Department of Materials Science \& Engineering, Northwestern University, Evanston IL 60208 USA
}%
\affiliation{ 
Center for Computation and Theory of Soft Materials, Northwestern University, Evanston IL 60208 USA
}%
\affiliation{Department of Chemistry, Northwestern University, Evanston IL 60208 USA.}
\affiliation{Department of Chemistry, Northwestern University, Evanston IL 60208 USA.
Department of Physics \& Astronomy, Northwestern University, Evanston IL 60208 USA}
\date{\today}

\begin{abstract}
A body placed in a rigidly-rotating fluid becomes circumscribed by a fictitious cylinder with generators parallel to the axis of rotation, a Taylor column. Slowly-moving liquid impinging on the body will swerve around the cylinder. Thus, Taylor columns may form when a breeze impinges on a mountain or when slowly-moving oceanic water impinges on a seamount, both due to the Earth's rotation. 
Here we show that classical non-rotating 
liquids endowed with an odd or Hall coefficient of viscosity, exhibiting nondissipative behavior, also give rise to Taylor column structures resembling halos or spears. 
Steady three-dimensional flow of such a liquid becomes
effectively two dimensional, swirling around the Taylor column
imitating its rigidly-rotating 
counterparts. Formation of Taylor halos and spears is attributed to the propagation of data along characteristics that may be parallel or oblique to a center axis, respectively.  
\end{abstract}

\maketitle

{W}hen the Coriolis force dominates over inertia and shear viscous forces, 
slow flow incident on a stationary obstacle placed in a rigidly-rotating liquid is accompanied by a column of fluid circumscribing the body, having generators parallel to the axis
of rotation. These are the Taylor columns \cite{Batchelor1967}. 
Different
flow patterns form in the exterior and interior of the Taylor
column and
it is shear viscosity that will trigger communication between these regions 
\cite{moore1968}. Taylor columns are known in the geophysical and atmospheric sciences, but are 
also important to ecology. For instance, cold water and low
salinity domes forming over seamounts (eg. the Rockall, Faroe and 
Hutton Banks) are bestowed with high chlorophyll and nutrient levels enabling larval diversity hotspots \cite{Dransfeld2009}. The importance of the columns to living systems was appreciated by 
G. I. Taylor who observed that fish 
demonstrated awareness 
of the cylindrical surface surrounding the column, by trying to avoid it, as they try to avoid a solid obstacle \cite{Yih1988}. 


Taylor columns
are less well-understood in active matter. Bush \emph{et} al. \cite{Bush1994} have suggested that 
individual particles in suspensions are accompanied by their own Taylor columns and that 
it remains unclear how the latter affect hydrodynamic interactions. 

Soni \emph{et} al.\cite{Soni2019} showed
that odd viscosity is important for active matter. In particular, blobs suspended with micron-size spinning magnets showed an attenuation of surface undulations that could not be explained by the existing theory based on the Navier-Stokes equations endowed with shear and couple stresses. Incorporation of an odd viscous stress, however, brought consistency between theory and the experimental observations.  


\begin{figure}
\vspace{-18pt}
\centering
\includegraphics[width=.97\linewidth]{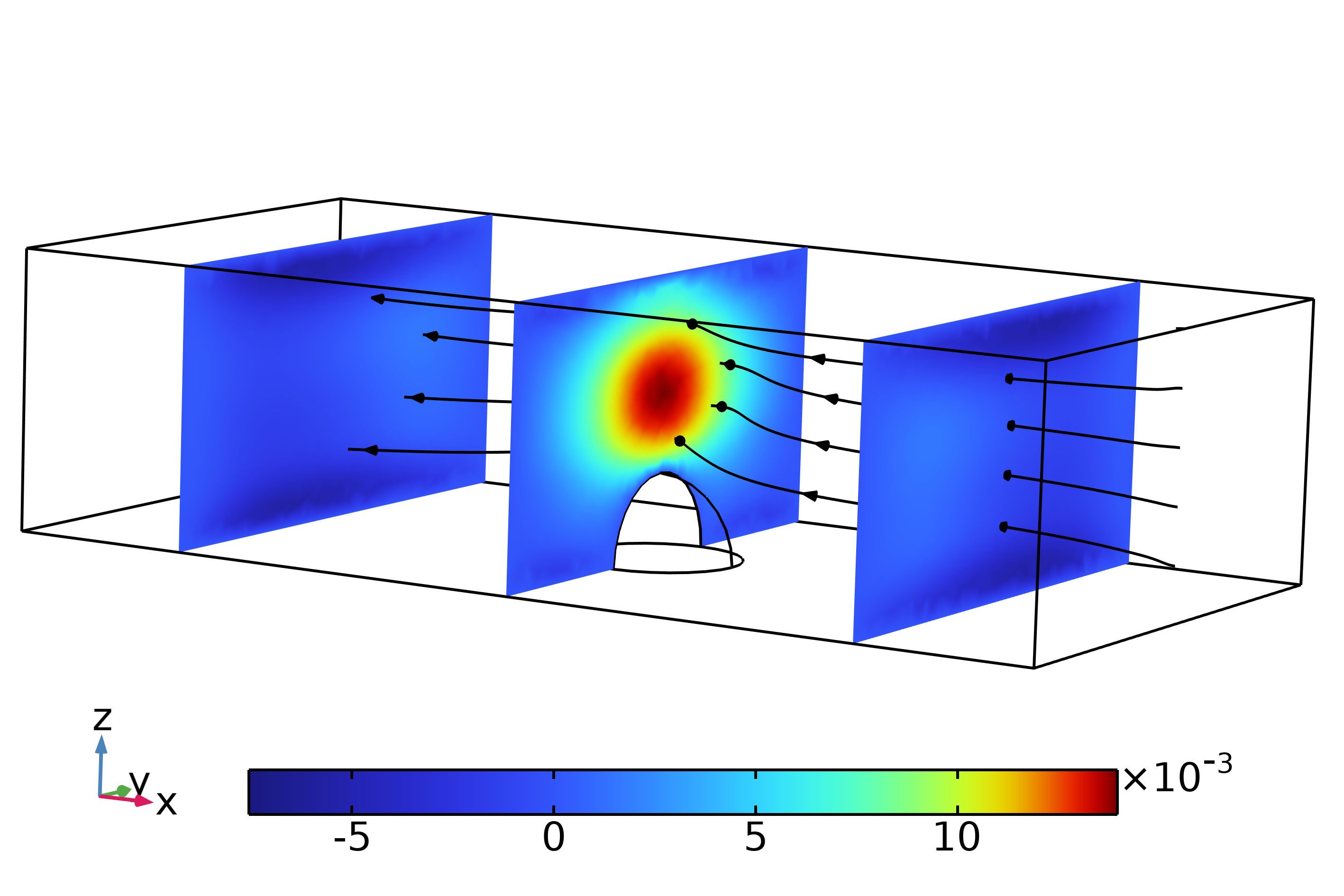}
\vspace{-5pt}
\caption{Odd viscous liquid entering from the right into a rectangular channel with velocity $u\hat{\mathbf{x}}$ encounters a mountain (a hemisphere)
on the ground, giving rise to a Taylor column
denoted by the sharp yellow and red colors. 
The colorbar shows transversal velocity $v\hat{\mathbf{y}}$. Streamlines (black curves) try to avoid entering the Taylor column
and swerve around it, as they also do in a rigidly-rotating liquid.}
\label{fig1}
\vspace{-5pt}
\end{figure}

In this article we point-out that 
liquids endowed with
odd viscosity, impinging on an
obstacle,
can form Taylor columns. In such three-dimensional liquids isotropy is no
longer present and an anisotropy axis exists, generated by a 
background mechanism such as a magnetic field or rotation. We assume that
such an axis has already been established and 
investigate the consequences of odd viscosity on fluid flow. 
Our theoretical discussion and numerical simulations proceed by assuming that odd viscosity dominates over its shear viscosity counterpart. This can be realized, for instance, by employing a liquid with a low
shear viscosity coefficient. 
Taylor columns can form when an obstacle moves slowly in a cylinder along the axis of anisotropy \cite{Kirkinis2023axial}. Here we consider a liquid impinging on an obstacle in a direction perpendicular to the anisotropy axis. 

Avron \cite{Avron1998} showed that a classical liquid with broken time-reversal symmetry (for instance due to a magnetic field or rotation) is endowed with a separate viscosity coefficient, called odd or Hall viscosity. A general expression of the resulting stress tensor 
can be cast in the form
\be \label{sigmaab}
\sigma_{\alpha\beta}' = \eta_{\alpha\beta\gamma\delta}D_{\gamma\delta},
\ee
where $D_{\gamma\delta} = \frac{1}{2}\left( \frac{\partial u_\gamma}{\partial x_\delta} + \frac{\partial u_\delta}{\partial x_\gamma}\right)$ is the rate-of-strain tensor.  
The viscosity tensor $\eta_{\alpha\beta\gamma\delta}$ is a collection of kinetic coefficients and thus
satisfies certain symmetries according to Onsager's principle \cite{Landau1980, Landau1981}.
When these 
coefficients depend on external fields, say $\mathbf{b}$, that 
change sign under time-reversal, the symmetry of the kinetic
coefficients is ensured when
\be
\eta_{\alpha\beta\gamma\delta}(\mathbf{b})
= \eta_{\gamma\delta\alpha\beta}(-\mathbf{b}).
\ee
The resulting stress tensor becomes \cite{Landau1981}
\begin{align} \nonumber
\sigma_{\alpha\beta}'& = \eta_o\left[ (b_{\alpha\gamma} b_\beta b_\delta + b_{\beta\gamma} b_\alpha b_\delta) D_{\gamma\delta}-b_{\beta\gamma}D_{\alpha\gamma} - b_{\alpha\gamma}D_{\beta\gamma} \right], \\
 &-2\eta_4(b_{\alpha\gamma} b_\beta b_\delta + b_{\beta\gamma} b_\alpha b_\delta)D_{\gamma\delta}\label{stress2}
\end{align}
where $b_{\alpha\beta} = \epsilon_{\alpha\beta\gamma}b_\gamma$, $\epsilon_{\alpha\beta\gamma}$ is the alternating tensor and the Einstein summation convention is implied on 
repeated indices. In \rr{stress2} we introduced a stress tensor involving two (odd) viscosity coefficients. In a scientific field that for a long time studies were concentrated on fluid flow behavior due to a single (shear) viscosity coefficient, it is advisable that we study the effects of $\eta_o$ and $\eta_4$
in turn. Setting $\eta_4=0$,
and considering
the anisotropy axis $\mathbf{b}$ to point, for instance, in the $z$-direction
the odd stress tensor \rr{stress2} obtains the form 
\be \label{sigma1}
\bm{\sigma}' = \eta_o 
\left(\begin{array}{ccc}
-\left(\partial_x v + \partial_y u \right) &  \partial_x u  - \partial_y v & 0\\
 \partial_x u  - \partial_y v & \partial_x v + \partial_y u   & 0\\
0&0&0
\end{array}
\right),
\ee
$\eta_o$ is the odd or Hall viscosity coefficient and $\mathbf{u} = (u,v,w)$ the 
three-dimensional liquid velocity field in Cartesian coordinates. 
Note that this form of the odd stress tensor
can not be simplified as is the case in two dimensions \cite{Ganeshan2017} since the flow here is three dimensional and
subject to the isochoric constraint
$\partial_x u + \partial_y v + \partial_z w =0$. It is clear that
the stress tensor \rr{sigma1} is symmetric and deviatoric (of zero trace). 
With \rr{sigma1}, the (odd) Navier-Stokes equations become
\be \label{NS}
\rho \frac{Du}{Dt} = -\frac{\partial p}{\partial x} - \eta_o\nabla^2_2 v, \;
\rho \frac{Dv}{Dt} = -\frac{\partial p}{\partial y} + \eta_o\nabla^2_2 u,\;
\rho \frac{Dw}{Dt} = -\frac{\partial p}{\partial z} 
\ee
where $D/Dt$ is the convective derivative,  $\nabla^2_2 = \partial_x^2 + \partial_y^2$ and $\rho$
the density of the liquid.
From \rr{NS} it can be observed that the 
odd viscous force density behaves like
a diffusive Coriolis force, where the liquid 
rotates with different angular velocity at different length scales of the lateral plane (the plane normal to the anisotropy axis) and is independent of elevation.
This observation implies that
under certain conditions the 
flow is expected to effectively
be two-dimensional, as 
is the case in rigidly rotating
liquids subject to the Taylor-Proudman theorem. This theorem is briefly summarized below and 
adopted to describe steady motions of an odd viscous liquid. 


Consider an incompressible inviscid liquid rotating rigidly about the $z$-axis with angular velocity $\Omega$ in the absence of odd viscosity. 
The geostrophic equations
\be \label{TP0}
\frac{\partial p}{\partial x} = -\rho\Omega v, \quad \frac{\partial p}{\partial y} = \rho\Omega u, \quad \frac{\partial p}{\partial z} = 0,  
\ee
show that the pressure $p$ is a streamfunction, it does not depend on $z$ and the
two components $u$ and $v$ of the velocity are likewise independent of $z$. 
Eliminating the pressure from Eq. \rr{TP0}
leads to $\frac{\partial u}{\partial x} + \frac{\partial v}{\partial y} =0$ and thus to 
$\partial_zw =0$ by employing the incompressibility condition. 
Summarizing, we found
\be \label{TP}
\partial_zu=\partial_zv = \partial_zw =0.
\ee
This is the Taylor-Proudman theorem stating that variations
of velocity along the rotation axis are decoupled from motions in
the lateral plane (the plane normal to the rotation axis), thus the flow effectively becomes two-dimensional.

\begin{figure}
\vspace{-30pt}
\centering
\includegraphics[width=.97\linewidth]{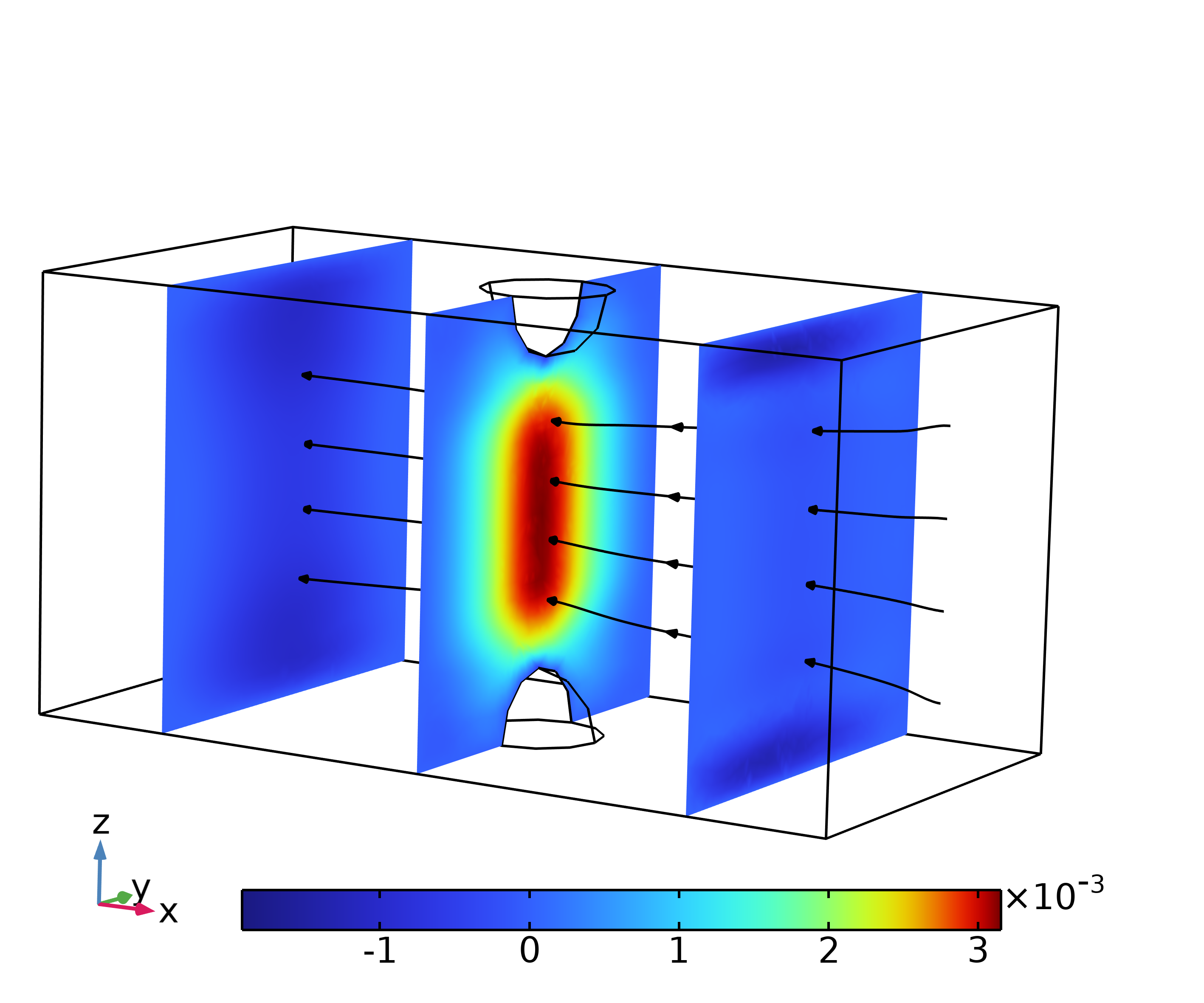}
\caption{Odd viscous liquid entering from the right into a rectangular channel with velocity $u\hat{\mathbf{x}}$ encounters \emph{two} obstacles (hemispheres)
one on
the ground and one on the ceiling, giving rise to a \emph{shared} Taylor column
denoted by the sharp yellow and red colors.
The colorbar shows strength of the transversal velocity $v\hat{\mathbf{y}}$. Streamlines (black curves) try to avoid entering the Taylor column as they also do in a rigidly-rotating liquid.}
\label{fig2}
\end{figure}

Condition $\frac{\partial w}{\partial z} = 0$ in \rr{TP} 
implies that when liquid flowing perpendicular
to the rotation axis, impinges on a solid body and is thus deflected, 
this deflection necessarily takes place in the same lateral plane and around the body
(otherwise there would be a variation of $w$ with respect to $z$).
Thus, a solid body in rotating flow will be accompanied by 
a column of liquid extending above, below and circumscribing
the body. Experiments show that dye released at the exterior,
never enters the column and vice-versa for the dye released in the
interior of the 
column \cite[Fig.16.2]{Tritton1988}.



The odd Navier-Stokes equations 
\rr{NS} can acquire a structure that is formally analogous
to the geostrophic equations \rr{TP0}
\be \label{TP1}
\frac{\partial p}{\partial x} = -\eta_o \nabla^2_2 v, \quad \frac{\partial p}{\partial y} = \eta_o \nabla^2_2 u, \quad \frac{\partial p}{\partial z} = 0. 
\ee
This reduction is possible by invoking the 
requirement $u \ll \nu_o /\ell$ where $u$ and $\ell$ are characteristic velocity
and length scales respectively, and
where $\nu_o = \eta_o/\rho$ is the liquid's odd kinematic viscosity.
This inequality can be derived by balancing the inertial terms
$\mathbf{u}\cdot \nabla \mathbf{u} \sim u^2/\ell$ with the odd viscous term $\nu_o \nabla_2^2 u\sim \nu_o u/\ell^2$ and requiring the latter to be dominant. 
\begin{figure}
\vspace{-10pt}
\centering
\includegraphics[width=.97\linewidth]{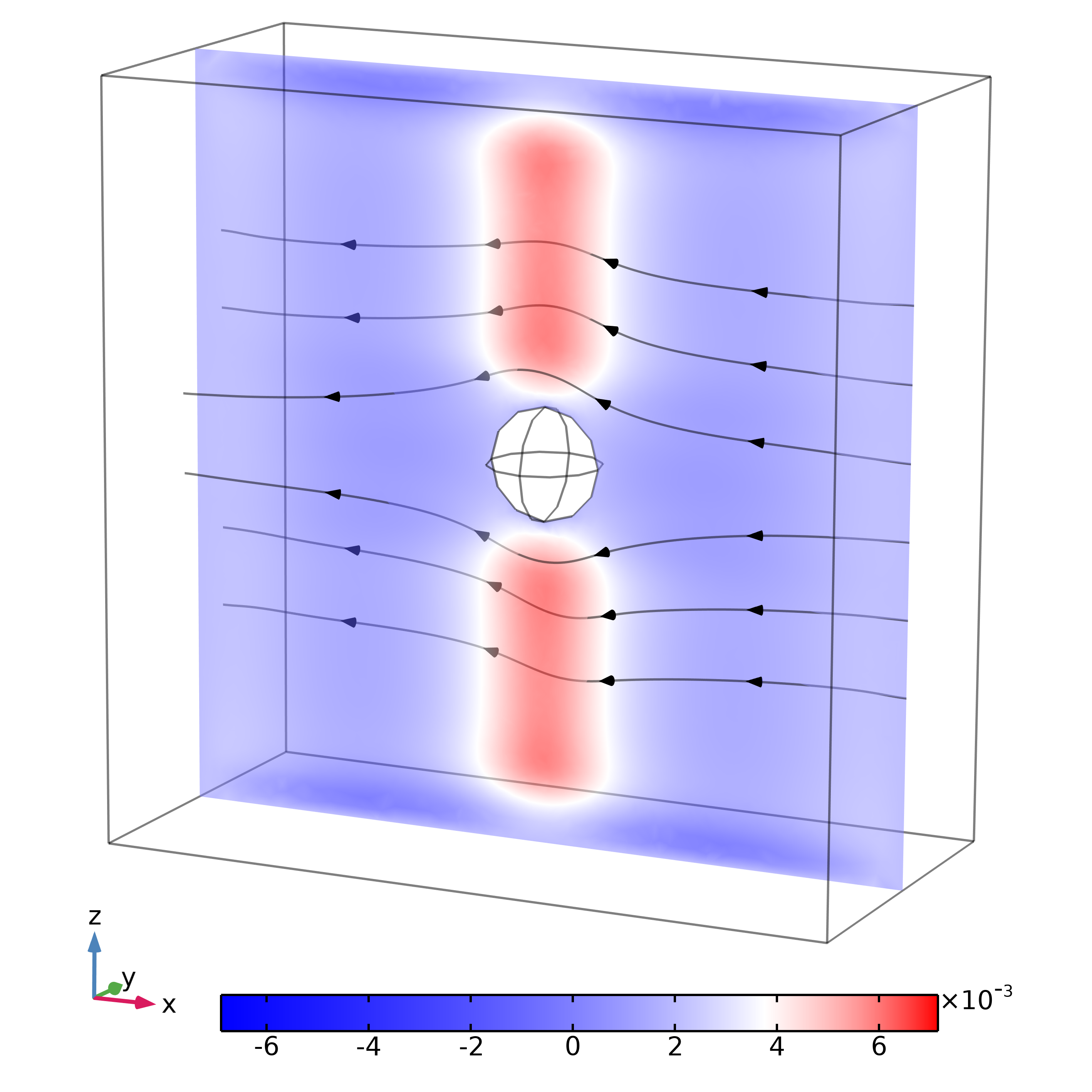}
\caption{Odd viscous liquid entering from the right into a rectangular channel with velocity $u\hat{\mathbf{x}}$ encounters a centered spherical obstacle giving rise to \emph{two} Taylor columns
denoted by the sharp red and white colors.
The colorbar shows strength of the transversal velocity $v\hat{\mathbf{y}}$. Streamlines (black curves) try to avoid entering the Taylor column as they also do in a rigidly-rotating liquid.}
\label{fig3}
\end{figure}
Introducing new velocity fields $(U,V,W)$ through
the substitution 
$
(U,V, W) = \ell^2\nabla^2_2(u,v,w),
$
(the length-scale is determined, for instance from the size of the vessel) makes the analogy between
\rr{TP0} and \rr{TP1} clear.
Thus, the following modified Taylor-Proudman theorem for the velocity field $U,V,W$ can be derived
\be \label{mTP}
\partial_z U = \partial_z V = \partial_zW =0,
\ee
and $\partial_xU + \partial_yV = 0$. 
There is a superposition of a two-dimensional
motion in the lateral ($x$-$y$) plane and a vertical motion, independent of $z$ for the velocity field
$(U,V,W)$ in analogy to the Taylor-Proudman theorem \rr{TP}.

Some familiar behavior at a boundary can be recovered. Because of the no-penetration
condition $w=0$ on a solid boundary we have $W = 0$ on the same boundary (since $(\partial_x^2 + \partial_y^2)w$ must be zero
on the boundary). Thus, when 
a streamline parallel to the axis meets a stationary boundary, this implies that $W$ is zero everywhere. 

The liquid satisfying Eq. \rr{mTP} can be considered as sitting in a frame rotating with angular velocity $\Omega = \nu_o\ell^{-2}$.
As is the case in rotating liquids, the force density associated
with the odd stress tensor \rr{sigma1} does not depend on the 
actual location of the anisotropy axis. Thus, the effects related
to odd viscosity are a consequence of the presence of the 
background "vorticity" $2\Omega = 2\nu_o\ell^{-2}$. 


Geostrophic flows are usually characterized by two dimensionless
quantities, the Rossby number and the inverse of the Taylor number 
(the Ekman number). Here, we can likewise define analogous dimensionless
numbers. These are the 
Taylor $\mathcal{T}$ and Maxworthy $\mathcal{M}$ numbers, the latter honoring one of the pioneers of experimental studies of Taylor columns \cite{Maxworthy1970},
\be
\mathcal{T} = \frac{\nu_o}{\nu_e}, \quad \mathcal{M} = \frac{\nu_o}{aU},  
\ee
where $a$ is the size of the obstacle, $U$ the impinging velocity and $\nu_e$ the kinematic shear viscosity. 
The Taylor number can be understood as an inverse Ekman number denoting the strength of odd to even viscosity 
and the Maxworthy number is the 
ratio of odd viscous force to inertial forces.
Thus, the conditions on the Navier-Stokes equations that lead to the modified Taylor-Proudman theorem
\rr{mTP} can be rephrased as $\mathcal{T}\gg 1$ and $\mathcal{M}\gg 1$. 

Figures \ref{fig1}-\ref{fig3}
display the numerically determined salient features of an odd viscous liquid entering a rectangular box with no-slip walls and encountering one or two no-slip obstacles. 
Streamlines swerve around the Taylor column. Within the column, a sharp gradient 
of transversal velocity forms.
It is due to
the emission of inertial waves
from the body, carrying the 
information of its presence
until they meet the solid wall
and are reflected back, thus creating a standing wave pattern \cite{Davidson2013book}. Transversal velocities are known
to form inside Taylor columns
in an axisymmetric geometry
\cite{Maxworthy1970}. In these 
figures the Taylor number is 
$\mathcal{T} = 50 $, the Maxworthy number $\mathcal{M} = 26 $ and the
shear viscosity is that of water.
In these numerically-determined flow patterns we do observe deviations from the Taylor-Proudman theorem associated with 
a finite value of shear viscosity. 

At small length scales $\ell$
where
$\mathcal{M}\ll \frac{\ell}{a} <1$ and
$a$ is the obstacle radius,
viscous effects in thin layers may become 
prominent. 
Boundary layers will form between the liquid and a solid surface as well as 
between the exterior and interior of the Taylor column. For a detailed study of these boundary layers in rigidly rotating liquids the reader can refer to 
\cite{moore1968, Bush1994,tanzosh1994}.

Odd viscosity is also responsible for modifying vortex twisting
and vortex stretching, cf. \cite[\S 6.6]{Tritton1988}. 
The derivatives $\partial_z U$ and $\partial_z V$ are intimately related to vortex twisting: nonzero values imply that fluid flow approaching a solid body
at two different elevations will have different values of these
derivatives thus generating nonzero vorticity in the plane perpendicular
to the rotating axis. Similarly, nonzero values of $\partial_z W$
imply that vorticity is modified parallel to the rotation axis: 
None of these statements are in effect when the conditions that validate
the Taylor-Proudman theorem hold.

We now return to the stress tensor \rr{stress2} and consider the effect 
the odd viscosity coefficient $\eta_4$
has on fluid-flow. Setting $\eta_o=0$
in \rr{stress2} and considering
the anisotropy axis $\mathbf{b}$ to point, again, in the $z$-direction
we obtain
\be \label{sigma4}
\bm{\sigma}' = \eta_4 
\left(\begin{array}{ccc}
0 & 0  & -(\partial_y w + \partial_z v)\\
0 & 0  & \partial_x w + \partial_z u\\
 -(\partial_y w + \partial_z v)&\partial_x w + \partial_z u&0
\end{array}
\right), 
\ee
cf. \cite[Eq. (58.16)]{Landau1981}. 

The foregoing construction of the Taylor-Proudman
theorem can be repeated by combining both odd viscous stress coefficients $\eta_o$ and $\eta_4$. We thus obtain 
the three partial differential equations 
\rr{mTP} where now the three "velocity" fields are defined
by $
(U,V, W) = \mathcal{S}(u,v,w),
$
so that 
\be \label{mTP4}
\partial_z\mathcal{L}u=0, \quad 
\partial_z\mathcal{L}v=0, \quad 
\partial_z\mathcal{L}w=0,
\ee
where 
\be \label{Leta4}
\mathcal{S} = (\nu_o-\nu_4)\nabla^2_2 + \nu_4\partial_z^2, 
\ee
and $\nu_4 = \eta_4/\rho$. 

\begin{figure}
\vspace{-10pt}
\centering
\includegraphics[width=.97\linewidth]{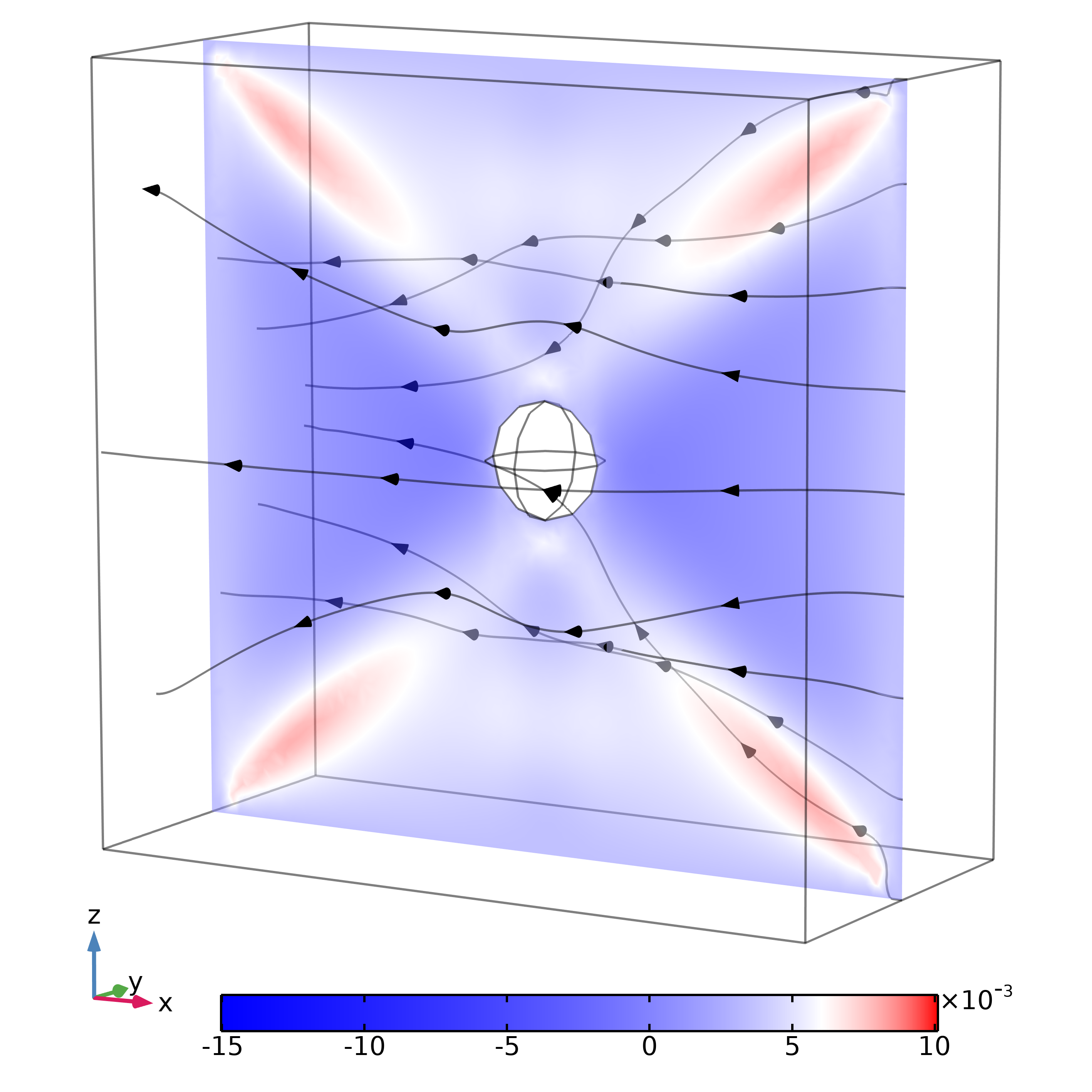}
\caption{Odd viscous liquid characterized by odd viscosity coefficient $\eta_4$, see Eq. \rr{sigma4}(setting $\eta_o =0$) entering from the right into a rectangular channel with velocity $u\hat{\mathbf{x}}$ encounters a centered spherical obstacle giving rise to \emph{four} Taylor spears
denoted by the sharp red and white colors.
The colorbar shows strength of the transversal velocity $v\hat{\mathbf{y}}$. Streamlines (black curves) try to avoid entering the Taylor spears and, in general, they follow the characteristic curves $x=\pm z$.}
\label{fig4}
\end{figure}
In Fig. \rr{fig4} we display the transversal velocity
$v\hat{\mathbf{y}}$
where now $\nu_0=0$ and 
$\nu_4 = 0.1 \textrm{cm}^2/$sec. Apparently there are Taylor column structures (spears) appearing at a $45$ degree angle with respect to the vertical. This behavior can be easily understood by analyzing each of the PDEs
in \rr{mTP4} whose meaning is determined by the form of the differential operator 
\rr{Leta4}. For instance, the
first, $\partial_z\mathcal{L}u=0$, can be written in the form 
\be \label{xyzpde}
\partial_z(\partial_x - \partial_z)(\partial_x+\partial_z) u =0.
\ee
Assuming that $z$ plays the role of the time-like variable, 
as is standard practice in the 
rotating liquid literature
\cite{Landau1987, Whitham1974},
(for simplicity we consider a slice so we set $\partial_y=0$), 
the characteristic equations associated 
with \rr{xyzpde}
read 
\be
\frac{dx}{dz} = \pm 1, \quad \frac{du}{dz}=0.
\ee
Thus, "initial" data propagate along the characteristic directions $x=\pm z$
\cite{Kevorkian2000}. On the other hand, the Taylor column displayed in Fig. \rr{fig3}, characteristics are lines along the $z$ direction and data can only propagate vertically. 
The streamline pattern displayed in Fig. \ref{fig4} by the black curves also shows this tendency to move along the characteristics $x=\pm z$.


Taylor-like columns are known to form in quasi-steady rotating turbulence \cite{Davidson2013book}. Odd viscosity will then compete with rotation to inhibit or assist the formation
of these columns and to generate (finite-amplitude) inertial waves. In material science, instabilities in the form of distorted shapes develop during 
nanowire growth \cite{schwalbach2012}. Odd hydrodynamics
described in this article,
may suppress the instability by 
wrapping each nanowire in its 
own Taylor column, thus providing 
a unidirectional wave-guide for their growth.

\begin{acknowledgments}
This research was supported by the US Department of Energy, Office of Science, Basic Energy Sciences under Award No. DE-SC0000989.
\end{acknowledgments}

\section*{Data Availability Statement}

The data that support the findings of this study are available from the corresponding author upon reasonable request.

\bibliography{disorder,fluids,materialscience, perturbations}

\end{document}